\documentclass[11pt,a4paper]{article}
\pdfoutput=1

\usepackage{jheppub}
\usepackage{latexsym}
\usepackage{multirow}
\usepackage{color}
\usepackage[usenames,dvipsnames,table]{xcolor}
\usepackage{graphicx}
\usepackage{epsfig}  
\usepackage{epsf}    
\usepackage{dcolumn}
\usepackage{bm}
\usepackage{dcolumn}
\usepackage{textcomp}
\usepackage{float}
\usepackage{subfig}
\usepackage{hypcap}
\usepackage[]{hyperref}
\usepackage{makecell}
\usepackage{color}
\usepackage{pifont}
\usepackage{appendix}
\usepackage{url}
\usepackage{cancel}

\hypersetup{
  bookmarks=true,         
  unicode=false,          
  pdftoolbar=true,        
 pdfmenubar=true,        
 pdffitwindow=true,     
 pdfstartview={FitH},    
 pdfsubject={Neutrino Oscillations Phenomenology},   
 pdfnewwindow=true,      
 pdfcreator={RevTeX},
 colorlinks=true,       
 linkcolor=red,          
 citecolor=blue,        
 filecolor=black,      
 urlcolor=blue,           
  }

\makeatletter
\renewcommand{\fnum@table}{\textbf{\tablename~\thetable}}
\renewcommand{\fnum@figure}{\textbf{\figurename~\thefigure}}
\makeatother

\preprint{TIFR/TH/17-04}

\title{Natural emergence of neutrino masses and dark matter  from $R$-symmetry} 

\author[a]{Sabyasachi Chakraborty,}
\author[b]{Joydeep Chakrabortty}

\affiliation[a]{Department of Theoretical Physics, Tata Institute of Fundamental Research, \\ 1, Homi Bhabha Road, Mumbai 400005, India}
\affiliation[b]{Department of Physics, Indian Institute of Technology, Kanpur-208016, India}

\emailAdd{sabya@theory.tifr.res.in}
\emailAdd{joydeep@iitk.ac.in}

\abstract
{We propose a supersymmetric extension of the Standard Model (SM) with a continuous global $U(1)_R$ symmetry. The $R$-charges of the SM fields are identified with that of their lepton numbers. As a result, both bilinear and trilinear `$R$-parity violating' (RPV) terms could be present at the superpotential. However, $R$-symmetry is not an exact symmetry as it is broken by supergravity effects. Hence, sneutrinos acquire a small vacuum expectation value in this framework. However, a  suitable  choice of basis ensures that the bilinear RPV terms can be completely rotated away from the superpotential and the scalar potential. On the other hand, the  trilinear terms play a very crucial role in generating neutrino masses and mixing at the tree level. This is noticeably different from the typical $R$-parity violating Minimal Supersymmetric Standard Model. Also, gravitino mass turns out to be the order parameter of $R$-breaking and is directly related to the neutrino mass. We show that such a gravitino, within the mass range $200~\text{keV}\lesssim m_{3/2}\lesssim 0.1~\text{GeV}$  can be an excellent dark matter candidate. Finally, we looked into the collider implications of our framework.}

 \arxivnumber{1701.04566}

\begin{document}
\maketitle
\toccontinuoustrue
\section{Introduction}
 Neutrino oscillation experiments have firmly established the existence of tiny non-zero masses of active neutrinos and non-trivial mixing~\cite{Esteban:2016qun} in the lepton sector. Since neutrinos are massless within the paradigm of Standard Model (SM), therefore, neutrino physics is a natural testing ground for physics beyond the SM (BSM). The most popular way to generate neutrino masses is through the see-saw mechanism. This also predicts Majorana nature of the neutrinos which signifies lepton number violation.  The basic idea behind the see-saw mechanism is to integrate out the heavy modes leading to higher-dimensional neutrino mass operators. Depending on the choice of the heavy particles one can classify variants of this mechanism, namely Type-I, -II, -III, Inverse, Double  see-saw {\it etc.,} both in supersymmetric (SUSY) and non-SUSY scenarios. Apart from neutrino masses and mixing, the deviation from the galactic rotation curve and bullet clusters provide concrete evidences in favor of dark matter (DM). Cosmological observations have also measured the relic density~\cite{Ade:2013zuv,Correia} of the DM with very high precision. But unfortunately the DM characteristics e.g. mass, spin and its nature i.e., cold, warm, single or multi component are yet to be determined. 

All these shortcomings of the SM can be explained quite efficiently in SUSY. For example, the lightest supersymmetric particle (LSP) is an excellent cold DM candidate in the paradigm of the minimal supersymmetric standard model (MSSM) with $R$-parity conservation. On the other hand, MSSM with $R$-parity violation (RPV) is an intrinsic SUSY way to generate neutrino masses and mixing both at the tree level as well as at the one-loop~\cite{Barbier:2004ez} level. However, if $R$-parity is broken, the LSP becomes unstable and hence fails to explain the observed relic density of the universe. In such cases, the DM candidate could be a gravitino, axino, axion and keV sterile neutrino~\cite{Feng:2010gw}.
 
Concerning the experimental verification of proposed models, unfortunately, the early 13 TeV run of the LHC has not found any signals~\cite{ATLAS:2016kts,CMS:2016mwj} in favor of SUSY. Although the non-observation of superpartners does not invalidate the idea of SUSY, it certainly questions the ability of MSSM to resolve the naturalness problem. So far, LHC has already ruled out gluinos lighter than 2 TeV when the gluino and LSP masses are well separated. This in turn makes the whole colored sector heavy due to the logarithmic sensitivity to the ultra-violet (UV) scale through renormalisation group evolutions. Interestingly, this correlation is not generic and can be avoided within the models of $R$-symmetry and Dirac gauginos. One needs to extend the gauge sector of MSSM to $N=2$ representation to  construct a Dirac gaugino mass. This require chiral superfields such as a  singlet $\widehat S$, a triplet $\widehat T$ under $SU(2)_L$ and an octet $\widehat O$ under $SU(3)_C$, in the adjoint representation of the SM gauge group. These fields couple with  bino, wino and gluinos respectively to generate Dirac masses for the gauginos. The presence of additional adjoint scalars cancel the UV logarithmic divergence for squark masses which results in a finite correction~\cite{Fox:2002bu} only. Hence, the Dirac gluino masses can easily be made heavy. An immediate consequence of having heavy Dirac gluinos is the suppression of the gluino pair or squark-gluino associated pair production cross-sections due to kinematic suppression. However, it is important to note that gluino pair production proceeds through QCD, and the production cross-section for Dirac gluinos would be twice as large compared to the Majorana gluinos with equal masses~\cite{diCortona:2016fsn}. This is based on the fact that Dirac gluinos have twice as many degrees of freedom than the Majorana gluinos. In addition, the pair production cross-section of squarks are also suppressed as it requires chirality flipping Majorana gluino masses in the propagator which is absent in these scenarios. This invariably  weakens~\cite{Kribs:2012gx} the stringent bound on the first two generation squark masses. Also, trilinear scalar couplings ($A$-terms) are absent in an $R$-symmetric framework and as a result flavor and CP violating interactions are suppressed~\cite{Kribs:2007ac}. 

Motivated by the issues pertaining to neutrino masses, DM and the non-observation of superpartners at the LHC, we propose a SUSY framework with an $R$-symmetry. Our prime aim in this paper is to generate neutrino masses within the $R$-symmetric Dirac gaugino framework. The $R$-charges are now identified with the lepton number of the SM fields. In general, we have the freedom to cast these charges and one can look into other assignments in \cite{U1R1,U1R2,U1R3,U1R4,U1R5,U1R6,U1R7,U1R8,Coloma:2016vod}. An outcome of our $R$-charge assignment is the presence of bilinear and trilinear ``$R$-parity violating'' terms in the superpotential which are $R$-symmetric. When  $R$-symmetry gets broken, the  bilinear RPV terms in the superpotential generate bilinear RPV terms in the scalar potential.  As a result, sneutrinos can acquire tiny vacuum expectation values (VEVs) proportional to the order parameter of $R$-breaking, i.e., gravitino mass. However, the superpotential and scalar potential have related sources of bilinear RPV. Therefore, such terms can be simultaneously rotated away. Nevertheless, the trilinear RPV terms will be present in the superpotential resulting in a mixing between light neutrinos and Dirac gauginos.  Hence, one generates light neutrino masses at the tree-level. The RPV couplings will also allow the gravitino to decay to a neutrino and a photon. Such a gravitino is an excellent decaying dark matter candidate provided its lifetime is greater than the age of the universe and is consistent with diffuse gamma ray fluxes.

We categorize our paper in the following manner. We propose the model and discuss the basic features of our scenario in section~\ref{sec:2}. We emphasize on the specific choice of $R$-charges leading to the presence of bilinear and trilinear ``RPV'' terms in the superpotential. In section~\ref{sec:3}, we discuss soft SUSY breaking, $R$-symmetry breaking and also the generation of sneutrino VEVs. Before proceeding to the fermionic sector, we choose a particular basis to remove the redundancy of the RPV operators in this model. Physics should be independent of the choice of basis. Therefore, we define two basis independent parameters $\kappa$ and $\zeta$ through which bilinear RPV is manifested. In section~\ref{subsec:basis}, we show that within the paradigm of our construction, the bilinear RPV terms can be rotated away completely from the superpotential and the scalar potential. However, the trilinear terms will remain. In section~\ref{sec:4}, we explore the neutral fermion mass matrix. There we illustrate how successfully neutrino masses and mixing can be generated in a simplistic scenario through the trilinear superpotential couplings. While considering normal as well as inverted hierarchies, we obtain the constraints on the relevant superpotential parameters. In section~\ref{sec:5} we explore the possibility of gravitino dark matter in our model satisfying necessary constraints.  We conclude by providing a direction to explore this scenario in collider experiments in section~\ref{sec:6}.

\section{\label{sec:2} The Model}
Our scenario is based on a SUSY framework with an added global $R$-symmetry where the gauginos are Dirac type unlike the MSSM. The choice of $R$-charges are shown in table~\ref{table:1}.
\begin{table}[h!]
\centering
 \begin{tabular}{ccc|ccc} 
 \hline
 Superfields & SM rep & $U(1)_R$ &  Superfields & SM rep & $U(1)_R$ \\ [0.5ex] 
 \hline\hline
 $\widehat H_u$ & $(1,2,1)$ & 0 & $\widehat R_d$ & $(1,2,-1)$ & 2 \\
 $\widehat H_d$ & $(1,2,-1)$ & 0 & $\widehat R_u$ & $(1,2,1)$ & 2\\
 \hline
 $\widehat Q_i$ & $(3,2,\frac{1}{3})$ & 1 & $\widehat S$ & $(1,1,0)$ & 0 \\ 
 $\widehat U_i^c$  & $(\bar 3,1,-\frac{4}{3})$ & 1 & $\widehat T$  & $(1,3,0)$ & 0 \\
 $\widehat D_i^c$ & $(\bar 3,1,\frac{2}{3})$ & 1 & $\widehat{\mathcal O}$ & $(8,1,0)$ & 0 \\
 \hline 
 $\widehat L_i$ & $(1,2,-1)$ & 2 \\
 $\widehat E_i^c$ & $(1,1,2)$ & 0 \\
 \hline
\end{tabular}
\caption{The gauge quantum numbers under the SM gauge group $SU(3)_c \bigotimes SU(2)_L \bigotimes U(1)_Y$ as well as the $U(1)_R$ charge assignments of the chiral superfields residing in the model.}
\label{table:1}
\end{table}
It is rather straightforward to show that the scalars share the same $R$-charges with their  corresponding superfields whereas for fermions they are one less. Similarly, the gauginos have $R$-charge one and the corresponding gauge bosons have  zero $R$-charge. From table~\ref{table:1}, we note that the lepton number of the SM particles can be identified with their $R$-charges. The lepton number of the superpartners can be non-standard. The definition of our $R$-charge assignment can also be understood as our choice of $R$ equals $R_0+L$~\cite{U1R3}. Here $R_0$ is the traditional $R$-charge assignment in MRSSM~\cite{RSYM1,RSYM2,RSYM3,RSYM4} {\footnote {We note in passing that $R_0$ is 1 for the lepton and quark superfields.}} and $L$ stands for the lepton number. In addition, an invariant action demands the superpotential to have $R$-charge of two units. Hence, the allowed terms in superpotential are:  
\begin{eqnarray}
W_{\text{MSSM}} &=& y_u^{ij}\widehat U_i^c\widehat Q_j\widehat H_u - y_d^{ij}\widehat D_i^c\widehat Q_j\widehat H_d - y_e^{ij}\widehat E_i^c\widehat L_j\widehat H_d-y_r^{i}\widehat E_i^c\widehat R_d\widehat H_d, \nonumber \\
W_{\text{adj}} &=& \lambda^u_S \widehat S\widehat R_d\widehat H_u-\lambda^u_T \widehat H_u\widehat T\widehat R_d+\lambda^d_S \widehat S\widehat H_d\widehat R_u-\lambda^d_T\widehat R_u\widehat T\widehat H_d, \nonumber \\
W_{\text{``bi-RPV"}} &=& \mu_i \widehat H_u\widehat L_i, \nonumber \\
W_{\text{``tri-RPV"}} &=& \xi_i \widehat S\widehat L_i\widehat H_u-\eta_i \widehat H_u\widehat T\widehat L_i, \nonumber\\
W_{\mu} &=& \mu_u\widehat R_d\widehat H_u+\mu_d\widehat H_d\widehat R_u.
\label{eq:superpot}
\end{eqnarray}
The total superpotential is then $W=W_{\text{MSSM}}+W_{\text{adj}}+W_{\text{``bi-RPV"}}+W_{\text{``tri-RPV"}} +W_{\mu}$. The triplet $\widehat T$ under $SU(2)_L$ is parametrised as
$\widehat T = \sum_{a=1,2,3}\widehat T^{(a)}, $ where $\widehat T^{(a)}=T_a \sigma^a/2$, $\sigma^a$'s being the Pauli matrices. We denote the components of the triplet field as $T_3=T_0$, $T_{+}=(T_1-iT_2)/\sqrt{2}$ and  $T_{-}=(T_1+iT_2)/\sqrt{2}$. In eq.~(\ref{eq:superpot}), $\widehat H_u, \widehat H_d, \widehat L_i, \widehat E_i^c, \widehat Q_i, \widehat U_i^c, \widehat D_i^c$ are the usual MSSM fields. Further, $\lambda^u, \lambda^d, \xi_i, \eta_i, y_u, y_d, y_e$ are trilinear/Yukawa couplings and $\mu_i$, $\mu_u$ and $\mu_d$ are couplings with mass dimension one. The traditional higgsino mass term ($\mu$) is forbidden in $R$-symmetric models.  To generate higgsino-like chargino and neutralino masses, it is mandatory to include two additional $SU(2)$-doublet chiral superfields $\widehat R_u$ and $\widehat R_d$ carrying non-zero $R$-charges. The presence of an $R$-charge for these two fields imply that $R$-symmetry cannot be spontaneously broken in the visible sector. Otherwise one has to encounter massless $R$-axions. These doublets are also known as inert doublets in the literature.

We like to stress that both the bilinear and trilinear ``RPV" but $R$-symmetric terms are present in the superpotential due to the assignment of $R$-charges. Also $R$-symmetry prohibits  baryonic ``RPV" terms in the superpotential and in the process the stringent constraints from proton decays can be circumvented. Before discussing neutrino mass generation mechanism, we would like to first address soft SUSY breaking, $R$-symmetry breaking and the generation of sneutrino VEVs. 

\subsection{Soft (super-soft) SUSY breaking interactions}
\label{subsec:1}
We choose to work in a scenario where SUSY (global) breaking is not associated with $R$-symmetry breaking. This can be achieved through  both $D$-- and $F$--type spurions. For example, Dirac gaugino masses can be  generated with the help of a spurion superfield $W_{\alpha}^{\prime}=\lambda_{\alpha}^{\prime}+\theta_{\alpha}D^{\prime}$ as~\cite{Benakli:2012cy,Goodsell:2015ura}:
\begin{eqnarray}
\mathcal L^{\text{gaugino}}_{\text{Dirac}} &=& \int d^2\theta \frac{W_{\alpha}^{\prime}}{\Lambda}\left[\kappa_1 W_{1\alpha}\widehat S+\kappa_2\text{Tr}(W_{2\alpha}\widehat T)+\kappa_3\text{Tr}(W_{3\alpha}\widehat O)\right]+\text{h.c.},
\end{eqnarray}
where $W_{i\alpha}$ and $W^{\prime}_{i\alpha}$ have $R$-charge 1, i.e.,  $R[\lambda_{i\alpha}]=R[\lambda_{i\alpha}^{\prime}]=1$ and $R[D^{\prime}]=0$. The integration over the Grassmann co-ordinates generates the  Dirac gaugino masses $M_i^D\sim\kappa_i\langle D^{\prime}\rangle/\Lambda$. Here, $i=1,2,3$ represent masses for $U(1)_Y$, $SU(2)_L$ and $SU(3)_C$ gauginos and $\Lambda$ refers to the messenger scale.
After the discovery of Higgs boson with mass around 125 GeV, it is important to address the status of Higgs mass within the given scenario. In a purely supersoft scenario, the scalar masses are one-loop suppressed compared to the gaugino masses. Consequently, because of $D$-flatness of the scalar potential~\cite{Fox:2002bu}, the tree level quartic term for the Higgs field is vanishingly small. This is challenging from the perspective of fitting the Higgs mass around 125 GeV. Thus instead of working in the generalized supersoft supersymmetry framework~\cite{Tuhin}, we  consider $F$-type breaking ~\cite{U1R3,U1R5} also. In our model, the scalar masses can be generated through such $F$-type spurion defined as $\widehat X=x+\theta^2 F_X$~\cite{U1R3} which allows the following $U(1)_R$ preserving operators
\begin{eqnarray}
&&\int d^4\theta \frac{\widehat X^{\dagger}\widehat X}{\Lambda^2}\Big[\sum_{i}\widehat\Phi_i^{\dagger}\widehat\Phi_i+\Big\{\widehat H_u \widehat H_d+\epsilon\Lambda\widehat S+\widehat S^2+\widehat T^2+\left(\frac{1}{\Lambda}\times\text{cubic}\right)+\text{h.c.}\Big\}\Big],\nonumber \\
&&\int d^2\theta \frac{X}{\Lambda}\left(\widehat S\widehat T^2+\widehat S\widehat O^2+\widehat S^3\right)+\text{h.c.},
\end{eqnarray}
and can automatically generate the following $U(1)_R$ preserving renormalizable soft SUSY breaking terms
\begin{eqnarray}
\mathcal L_{\text{soft}} &=& \sum_{i} m_i^2\phi_i^{\dagger}\phi_i+\left[t_S S+\frac{1}{2}b_S S^2+B_{\mu}H_u H_d+....\right].
\end{eqnarray}
The soft mass squared terms are proportional to $\langle F_X\rangle^2/\Lambda^2\equiv M_{\text{SUSY}}^2$ where we consider same magnitude for the $F$- and $D$-type VEVs. Such a mechanism also generates a scalar singlet tadpole $t_S S$. However, as long as $t_S < M_{\text{SUSY}}^3$, such a tadpole is not expected to destabilize the hierarchy~\cite{Goodsell:2012fm}. Nevertheless, due to the absence of $R$-breaking terms $B_{\mu_i}H_u \widetilde{\ell}_i$ in the scalar sector, sneutrinos cannot acquire VEV. This is an important ingredient for neutrino mass generation in traditional bilinear RPV scenarios. 

\subsection{\label{sec:3}{$R$-symmetry breaking}}

It is well established that our universe is associated with a vanishingly small vacuum energy or cosmological constant. To explain this from the perspective of spontaneously broken supergravity theory in the hidden sector, the superpotential needs to acquire a non-zero VEV. Since the superpotential carries non-zero $R$-charge, therefore, $\langle W\rangle\neq 0$ implies breaking of $R$-symmetry. As a result, the gravitino would also acquire a mass which turns out to be the order parameter of $R$-symmetry breaking. The $R$-breaking information  is then communicated to the visible sector through anomaly mediation and in the process the following $R$-symmetry breaking terms are generated
\begin{eqnarray}
\mathcal L_{\cancel{R}}&=& M_1 \widetilde b\widetilde b
+ M_2 \widetilde w\widetilde w 
+ M_3 \widetilde g\widetilde g 
+ A_u \widetilde u_R \widetilde u_L^{\ast} H_u^0 
+ A_d \widetilde d_R \widetilde d_L^{\ast} H_d^0 
+ A_\ell \widetilde \ell_R \widetilde \ell_L^{\ast} H_d^0 + {\mathrm h.c.},
\label{eqn:Rbreak}
\end{eqnarray}
where the Majorana gaugino masses are generated through small $R$-breaking effects as 
\begin{equation}
M_i = \frac{g_i^2}{16\pi^2} b_i m_{3/2} ~~~(i = 1,2,3),
\end{equation}
with beta functions
\begin{equation}
b_1  = 33/5 , \qquad\qquad  b_2=1 , \qquad\qquad b_3=-3 \ .
\end{equation}
The small $R$~symmetry-breaking trilinear scalar interactions are as follows
\begin{eqnarray}
A_\tau &=& \frac{ m_{3/2}}{16\pi^2} 
\left( - \frac{9}{5}g_1^2 - 3g_2^2 + 3Y_b^2 + 4Y_\tau^2 \right),\nonumber  \\
A_t &=& \frac{ m_{3/2}}{16\pi^2} 
\left( - \frac{13}{15}g_1^2 - 3g_2^2 - \frac{16}{3}g_3^2 + 6Y_t^2 + Y_b^2 \right), \nonumber \\
A_b &=& \frac{ m_{3/2}}{16\pi^2} 
\left( - \frac{7}{15}g_1^2 - 3g_2^2 - \frac{16}{3}g_3^2 + Y_t^2 + 6Y_b^2 + Y_\tau^2 \right).
\end{eqnarray}

It is also important to note that the presence of a conformal compensator field $\Sigma = 1+\theta^2 m_{3/2}$ invariably generates a $B{\epsilon_i}$ term~\cite{Chacko:1999am}  in the superpotential through the following operator:
\begin{eqnarray}
\mathcal L =\int d^2\theta~\Sigma^3 \mu_i\widehat H_u\widehat L_i.
\label{Bep}
\end{eqnarray}
After scaling out this compensator field with $\widehat\Phi^{\prime}=\Sigma\widehat\Phi$ where $\widehat\Phi$ is a chiral superfield, we generate a bilinear term ($H_u\widetilde\ell_i$) in the scalar potential
\begin{eqnarray} 
B\mu_i=\mu_i m_{3/2}.
\end{eqnarray} 
Hence, the $B\mu$ term is always aligned with the $\mu$ term. Such terms are $R$-breaking effects and proportional to the gravitino mass. The presence of this small effect would generate tiny sneutrino VEVs which might become important for neutrino mass--mixing as we discuss in the next section.
\subsection{Sneutrino VEV}
\label{subsec:2}
To compute the sneutrino VEVs, one has to include  the contributions from $F$-, 
$D$- and soft SUSY breaking terms. The  additional pieces associated with $SU(2)_L$ and $U(1)_Y$ in the $D$-terms are
\begin{eqnarray}
D_2^a &=& g\left(H_u^{\dagger}\tau^a H_u+\widetilde\ell_i^{\dagger}\tau^a\widetilde\ell_i+T^{\dagger}\lambda^a T\right)+\sqrt{2}\left(M_2^D T^a+M_2^D T^{a\dagger}\right),
\end{eqnarray}
where $\tau^a$ and $\lambda^a$'s represent the $SU(2)$ generators in the fundamental and adjoint representations respectively. Similarly, the weak hyper-charge contribution $D_Y$ is given by
\begin{eqnarray}
D_Y &=& \frac{g^{\prime}}{2}\left(H_u^{\dagger}H_u-\widetilde\ell_i^{\dagger}\widetilde\ell_i\right)+\sqrt{2}M_1^D\left(S+S^{\dagger}\right).
\end{eqnarray}
The tree level scalar potential terms which participate in the sneutrino field minimization equations are
\begin{eqnarray}
V_F &=& |\mu_i|^2|\widetilde\nu_i^0|^2, \nonumber \\
V_{\text{soft}} &=& \widetilde m^2 |\widetilde\nu_i^0|^2+B\mu_i H_u^0\widetilde\nu_i, \nonumber \\
V_D &=& \left[\frac{(g^2+g^{\prime 2})}{8}|\widetilde\nu_i^0|^2-\sqrt{2}g^{\prime} M_1^D v_S +\sqrt{2}g M_2^D v_T \right]|\widetilde\nu_i^0|^2.
\end{eqnarray}
In the limit $v_S, v_T\rightarrow 0$, the sneutrino VEVs can be well approximated as
\begin{eqnarray}
\langle\widetilde\nu_i\rangle &=& -\frac{B\mu_i v_u}{\widetilde m_i^2+\mu_i^2}.
\label{eq:snvev}
\end{eqnarray}
Such a choice of the singlet and  triplet VEVs also ensure that these fields are very heavy through their respective minimization equations. Assuming $\langle
H_u^0\rangle=v_u\sim\widetilde m_i$ i.e., at the electroweak scale, we find $\langle\widetilde\nu_i\rangle\sim B\mu_i/\widetilde m\sim \mu_i m_{3/2}/\widetilde m$. Off course, in the same manner the inert scalars ($R_u, R_d$) would also acquire a VEV and as a result would mix with the Higgs fields. However, that mixing is also suppressed by the $R$-breaking parameter $m_{3/2}$ and does not play any important role in the phenomenological description. 
\subsection{Choice of basis}
\label{subsec:basis}
In the usual framework of bilinear RPV-MSSM~\cite{RPV2,RPV3,RPV5,Grossman:2003gq}, the lepton and the Higgs superfields are at the same footing~\cite{Grossman:1997is} as they carry the same gauge charges. The lepton number violating couplings depend on the choice of $(\widehat H_d, \widehat L_i)$ basis. Thus it is important to explicitly mention the choice of basis in which the analysis is being performed. However, physics should not depend on such choices.  Therefore, two basis independent  parameters $\sin\kappa$ and $\sin\zeta$ are often introduced in the literature~\cite{Banks:1995by,Bisset:1998bt,Binetruy:1997sm,Grossman:1998py} which encapsulate the total lepton number violation in the superpotential as well as in the scalar potential respectively.

In our scenario,  both the superfields $\widehat R_d$ and $\widehat L_i$ carry the same charges as can be seen from table~\ref{table:1}. Hence, in terms of the four-vector $\widehat L_{\alpha}$, $\alpha=0,1,2,3$ where $\widehat L_0\equiv \widehat R_d$,  the renormalizable superpotential can be written as:
\begin{eqnarray}
W &=& y_u^{ij}\widehat U_i^c\widehat Q_j\widehat H_u-y_d^{ij}\widehat D_i^c\widehat Q_j\widehat H_d-y_e^{i\alpha}\widehat E_i^c\widehat L_{\alpha}\widehat H_d +\xi_{\alpha}\widehat S\widehat L_{\alpha}\widehat H_u-\eta_{\alpha}\widehat H_u.\widehat T.\widehat L_{\alpha}+\lambda_s^d \widehat S\widehat H_d\widehat R_u\nonumber \\
&-&\lambda_T^d \widehat R_u\widehat T\widehat H_d 
+\mu_d \widehat H_d\widehat R_u +\mu_{\alpha}\widehat L_{\alpha}\widehat H_u.
\label{eq:sup_mod}
\end{eqnarray}
Similarly, the scalar potential consisting of soft and super-soft terms reduces to the following form:
\begin{eqnarray}
V_{\text{soft}} &=& \sum_i m_i^2\phi_i^{\dagger}\phi_i+ M_1^D\widetilde b\widetilde S+M_2^D \widetilde W\widetilde T+M_3^D \widetilde g\widetilde O+M_1 \widetilde b\widetilde b+M_2 \widetilde W\widetilde W+M_3 \widetilde g\widetilde g+B_{\alpha} H_u\widetilde L_{\alpha} \nonumber \\
&+&\left[t_S S+\frac{1}{2}b_S S^2+B_{\mu}H_u H_d+...\right]+A_u \widetilde u_R\widetilde u_L^* H_u^0+A_d \widetilde d_R\widetilde d_L^* H_d^0+A_l \widetilde l_R\widetilde l_L^* H_d^0.
\end{eqnarray}
In the zero-sneutrino VEV basis we define $\widehat R_d$ in terms of the $\widehat L_{\alpha}$ superfields as
\begin{eqnarray}
\widehat R_d &=& \frac{1}{v_R}\sum_{\alpha} v_{\alpha}\widehat L_{\alpha},
\label{eq:rd}
\end{eqnarray}
where $v_R\equiv \left(\sum_{\alpha} v_{\alpha} v^{\alpha} \right)^{1/2}$and gets generated due to $R$-symmetry breaking. Likewise, the four vector superfield $\widehat L_{\alpha}$ can now be defined in terms of the usual slepton superfields $\widehat L_i$ with vanishing VEVs and $\widehat R_d$ in the following way
\begin{eqnarray}
\widehat L_{\alpha} &=& \frac{v_{\alpha}}{v_R}\widehat R_d+\sum_i e_{\alpha i}\widehat L_i.
\label{eq:slepton}
\end{eqnarray}
Even then, there is a  freedom to rotate  the lepton ($\widehat L_i$) superfields arbitrarily. We choose that only a single lepton superfield couples to $\widehat H_u$ in the superpotential. This allows us to rewrite the superpotential in terms of basis independent quantities by plugging  eqs.~(\ref{eq:rd}) and~(\ref{eq:slepton}) in eq.~(\ref{eq:sup_mod}) as
\begin{eqnarray}
W &=& y_u^{ij}\widehat U_i^c\widehat Q_j\widehat H_u-y_d^{ij}\widehat D_i^c\widehat Q_j\widehat H_d-\widetilde y_e^i \widehat E_i^c\widehat R_d\widehat H_d-\widetilde y_e^{ij}\widehat E_i^c\widehat L_j\widehat H_d+\widetilde\lambda_S^u\widetilde S\widetilde R_d\widetilde H_u+\widetilde\xi_i\widehat S\widehat L_i\widehat H_u \nonumber \\
&-&\widetilde\lambda_T^u \widehat H_u\widehat T\widehat R_d-\widetilde\eta_i \widehat H_u\widehat T\widehat L_i +\lambda_S^d \widehat S\widehat H_d\widehat R_u-\lambda_T^d \widehat R_u\widehat T\widehat H_d+\mu_d \widehat H_d\widehat R_u \nonumber \\
&+&\mu_u\cos\kappa~\widehat R_d\widehat H_u+\mu_u\sin\kappa~\widehat L_3\widehat H_u,
\end{eqnarray}
where 
\begin{eqnarray}
\widetilde\xi_i &=& \sum_{\alpha} \xi^{\alpha}e_{\alpha i}, ~~~~~ \widetilde\lambda_S^u = \sum_{\alpha} \frac{v_{\alpha}}{v_R}\xi^{\alpha}, ~~~~~ \widetilde y_e^i = \sum_{\alpha}y_e^{\alpha i}\frac{v_{\alpha}}{v_R}, \nonumber \\
\widetilde\eta_i &=& \sum_{\alpha}\eta^{\alpha} e_{\alpha i}, ~~~~~\widetilde\lambda_T^u = \sum_{\alpha} \frac{v_{\alpha}}{v_R}\eta^{\alpha}, ~~~~~ \widetilde y^i_{e j}=\sum_{\alpha} y_e^{\alpha i}e_{\alpha j},
\end{eqnarray}
and 
\begin{eqnarray}
\cos\kappa \equiv \frac{1}{\mu_u v_R}\sum_{\alpha}\mu_{\alpha} v^{\alpha},
\end{eqnarray}
with $\mu_u\equiv\left(\sum_{\alpha}\mu_{\alpha} \mu^{\alpha} \right)^{1/2}$~\cite{Barbier:2004ez}. $\kappa$ is the angle between the four-vectors ${\mu}_{\alpha}$ and ${v}_{\alpha}$. 
It is evident from eq.~(\ref{eq:snvev}) that $v_{\alpha}\parallel B_{\alpha}$, i.e.,  $v_{\alpha}\parallel\mu_{\alpha}$ and therefore, $\sin\kappa=0$. Though $\mu_{\alpha}$ and $v_{\alpha}$  are basis dependent quantities their relative angle $\kappa$ does not depend on the choice of basis for $\widehat L_{\alpha}$ superfields. As a result, the effect of bilinear RPV terms can be rotated away completely from the superpotential.

Similarly, the bilinear $R$-parity violation in the scalar sector can be parametrised in terms of the angle $\zeta$ formed by four-vectors $B_{\alpha}$ and ${v}_{\alpha}$ as:
\begin{eqnarray}
\cos\zeta &\equiv& \frac{1}{B v_R}\sum_{\alpha}B_{\alpha} v^{\alpha},
\end{eqnarray}
where $B\equiv\left(\sum_{\alpha}B_{\alpha} B^{\alpha} \right)^{1/2}$.  But it is clear from the earlier section that $B_{\alpha},\;\mu_{\alpha},\; v_{\alpha}$ are also aligned together. This implies $\sin\zeta=0$ and thus allows us to rotate away the bilinear terms  from the scalar sector also. Therefore, in reality, the bilinear RPV terms do not  play any role in generating neutrino masses and mixing. However, trilinear terms in the superpotential can not be rotated away simultaneously. These terms play crucial role for neutrino mass generation as discussed in the following section.
\section{\label{sec:4}The fermion sector}
In this section we will consider both neutral  and charged fermion sectors. The mixing between neutral fermions and neutrinos lead to neutrino masses at the tree level.  Similarly, chargino mixes with the charged leptons which may potentially give  rise to lepton number violating processes. 
\subsection{The neutral fermion sector}
The Lagrangian corresponding to the neutral fermion sector after $R$-symmetry breaking contain the following terms
\begin{eqnarray}
\mathcal L_{f^0} &=& M_1^D \widetilde b\widetilde S+M_2^D \widetilde w^0\widetilde T+\mu_u\widetilde H_u^0\widetilde R_d^0+\mu_d\widetilde H_d^0 R_u^0+\widetilde\lambda^u_S v_u\widetilde S\widetilde R_d^0+\widetilde\lambda^d_S v_d\widetilde S\widetilde R_u^0+\widetilde\lambda^u_T v_u \widetilde T^0 \widetilde R_d^0 \nonumber \\
&+&\widetilde\lambda^d_T v_d\widetilde R_u^0\widetilde T^0 
+ M_1 \widetilde b\widetilde b+M_2 \widetilde w^0\widetilde w^0+M_S \widetilde S\widetilde S+M_T \widetilde T^0\widetilde T^0+\frac{g^{\prime} v_u}{\sqrt{2}}\widetilde b\widetilde H_u^0-\frac{g^{\prime} v_d}{\sqrt{2}}\widetilde b\widetilde H_d^0\nonumber \\
&-&\frac{g v_u}{\sqrt{2}}\widetilde w^0\widetilde H_u^0 
+\frac{g v_d}{\sqrt{2}}\widetilde w^0\widetilde H_d^0
+\widetilde\xi_{i}v_u\widetilde S\nu_i+\frac{\widetilde\eta_i v_u}{\sqrt{2}}\widetilde T^0\nu_i+\left(\widetilde\xi_i v_S+\widetilde\eta_i v_T\right)\widetilde H_u^0\nu_i.
\end{eqnarray}
Here, $``i"$ stands for $e$, $\mu$ and $\tau$. 

\noindent
The Lagrangian mass terms expressed in the basis $f^0\equiv( \widetilde b$,$\widetilde S$,$\widetilde w^0$,$\widetilde T^0$,$\widetilde H_u^0$,$\widetilde R_d^0$,$\widetilde H_d^0$,$\widetilde R_u^0, \nu_i)$ can be written schematically as
\begin{eqnarray}
\mathcal L_{f^0}^{\text{mass}} &=& \frac{1}{2}(f^0)^{T} M_{N} f^0,
\end{eqnarray}
where 
\begin{eqnarray}
M_{N} &=& \left(\begin{array}{cc}
m_{\widetilde f^0} & m_D \\
m_D^T & 0
\end{array}
\right),
\end{eqnarray}
with
\begin{eqnarray}
m_{\widetilde  f^0} &=& \left(\begin{array}{cccccccc}
M_1 & M_1^D & 0 & 0 & \frac{g^{\prime} v_u}{\sqrt{2}} & 0 & -\frac{g^{\prime} v_d}{\sqrt{2}} & 0 \\
M_1^D & M_S & 0 & 0 & 0 & \widetilde\lambda^u_S v_u & 0 & \widetilde\lambda_S^d v_d \\
0 & 0 & M_2 & M_2^D & -\frac{g v_u}{\sqrt{2}} & 0 & \frac{g v_d}{\sqrt{2}} & 0  \\
0 & 0 & M_2^D & M_T & 0 & \frac{\widetilde\lambda^u_T v_u}{\sqrt{2}} & 0 & \frac{\widetilde\lambda_T^d v_d}{\sqrt{2}} \\
\frac{g^{\prime} v_u}{\sqrt{2}} & 0 & -\frac{g v_u}{\sqrt{2}} & 0 & 0 & \mu & 0 & 0\\
0 & \widetilde\lambda^u_S v_u & 0 & \frac{\widetilde\lambda^u_T v_u}{\sqrt{2}} & \mu & 0 & 0 & 0 \\
-\frac{g^{\prime} v_d}{\sqrt{2}} & 0 & \frac{g v_d}{\sqrt{2}} & 0 & 0 & 0 & 0 & \mu \\
0 & \widetilde\lambda_S^d v_d & 0 & \frac{\widetilde\lambda_T^d v_d}{\sqrt{2}} & 0 & 0 & \mu & 0
\end{array}
\right),
\label{eq:neu_matrix}
\end{eqnarray}
and 
\begin{eqnarray}
m_D^T &=& \left(\begin{array}{cccccccc}
0 & \widetilde\xi_1 v_u & 0 & \frac{\widetilde\eta_1 v_u}{\sqrt{2}} & 0 & 0 & 0 & 0 \\
0 & \widetilde\xi_2 v_u & 0 & \frac{\widetilde\eta_2 v_u}{\sqrt{2}} & 0 & 0 & 0 & 0 \\
0 & \widetilde\xi_3 v_u & 0 & \frac{\widetilde\eta_3 v_u}{\sqrt{2}} & 0 & 0 & 0 & 0 
\end{array}
\right).
\end{eqnarray}
Here, $\langle H_u^0\rangle = v_u$, $\langle H_d^0\rangle = v_d$ with $v=\sqrt{v_u^2+v_d^2}$ and $\tan\beta=v_u/v_d$. We also consider $\mu_u\equiv\mu_d\equiv \mu$, and  $M_1\sim M_2\sim M_S\sim M_T\sim m_{3/2}/16\pi^2$. For simplicity, we have chosen $v_S$, $v_T\sim 0$. We note, an order one value of the superpotential couplings $\widetilde\lambda^u_{S,T}$ provide substantial one-loop corrections to the up-type Higgs~\cite{Diessner:2014ksa,Diessner:2015yna} boson mass $m^2_{H_u}$. These corrections are `stop-like' and an 125 GeV Higgs boson can be obtained without requiring too heavy top squarks. Hence, we kept $\widetilde\lambda^u_{S,T}$ but assumed $\widetilde\lambda^d_{S,T}$ to be small. In the next section, we carry out a simplified analysis to explain neutrino masses and mixings. For brevity, we shall also make the following transformations:
$\widetilde\lambda_S^u\sim\widetilde\lambda_T^u/\sqrt{2}\sim\lambda$, $\widetilde\eta/\sqrt{2}\to\widetilde\eta$.
\subsection{Neutrino mass and mixing}
\label{subsec:3}
In order to obtain a quantitative estimation of the relevant parameters which satisfy neutrino masses and mixing, we choose all the $R$-symmetry preserving masses are of the same order, i.e., $M_1^D\sim M_2^D\sim\mu\sim v_u\equiv \widetilde m$.  The structure of the effective neutrino mass matrix follows from the typical Type-I seesaw expression $[ m_D^T\;m_{\widetilde f^0}^{-1}\;m_D ]$ and represented as
\begin{eqnarray}
(m_{\nu})_{ij} \simeq \frac{m_{3/2}}{16\pi^2}\left[a~\widetilde\eta_i\widetilde\eta_j+\frac{b}{2}\left(\widetilde\eta_i\widetilde\xi_j+\widetilde\eta_j\widetilde\xi_i\right)+c~\widetilde\xi_i\widetilde\xi_j\right],
\label{eq:nu_mass}
\end{eqnarray}
where 
\begin{eqnarray}
a &=& -\frac{\left[2-2\sqrt{2}\lambda g^{\prime}+(g^2+g^{\prime 2})\lambda^2\right]}{\left[2+\lambda(g-g^{\prime})\{2\sqrt{2}+\lambda(g-g^{\prime})\}\right]}, \nonumber \\
b &=& -\frac{2\left[\sqrt{2}\lambda (g-g^{\prime})+(g^2+g^{\prime 2})\lambda^2\right]}{\left[2+\lambda(g-g^{\prime})\{2\sqrt{2}+\lambda(g-g^{\prime})\}\right]}, \nonumber \\
c &=& -\frac{\left[2+2\sqrt{2}\lambda g^{\prime}+(g^2+g^{\prime 2})\lambda^2\right]}{\left[2+\lambda(g-g^{\prime})\{2\sqrt{2}+\lambda(g-g^{\prime})\}\right]}. 
\label{eq:bterm}
\end{eqnarray}
In principle, the parameters $\widetilde\eta$ and $\widetilde\xi$ can be varied within the validated range to fit the observed neutrino masses and mixing as showed in~\cite{RPV1} for bilinear RPV scenario in MSSM. However, in that framework such a form of the neutrino mass matrix arises only after taking the loop corrections into account. Here we adopt a rather  simplified approach~\cite{RPV4} to estimate the values of these superpotential parameters such that neutrino masses and mixing can be successfully generated. The co-efficient of $\widetilde\eta_i\widetilde\xi_j$ vanishes for distinct values of $\lambda$ which can be obtained by setting $b=0$ in eq.~(\ref{eq:bterm}). Under such approximation, the neutrino mass expression in eq.~(\ref{eq:nu_mass}) turns out to be:
\begin{eqnarray}
(m_{\nu})_{ij}|_{\lambda=0} &\simeq& \frac{m_{3/2}}{16\pi^2}\left[\widetilde\eta_i\widetilde\eta_j+\widetilde\xi_i\widetilde\xi_j\right],\nonumber \\
(m_{\nu})_{ij}|_{\lambda=-\frac{\sqrt{2}(g-g^{\prime})}{(g^2+g^{\prime 2})}} &\simeq& \frac{m_{3/2}}{16\pi^2}\left[\frac{\left(1+\tan^2\theta_W\right)}{2}\widetilde\eta_i\widetilde\eta_j+\frac{\left(1+\cot^2\theta_W\right)}{2}\widetilde\xi_i\widetilde\xi_j\right].
\label{eq:check}
\end{eqnarray}
Further assuming $\widetilde\eta_i>\widetilde\xi_j$ ($\forall\; i,j$), we can decompose the full neutrino mass matrix as 
\begin{eqnarray}
(m_{\nu})_{ij}\simeq \frac{m_{3/2}}{16\pi^2}\left[c_0\underbrace{\left(\begin{array}{ccc}
\widetilde\eta_1^2 & \widetilde\eta_1\widetilde\eta_2 & \widetilde\eta_1\widetilde\eta_3 \\
\widetilde\eta_1\widetilde\eta_2 & \widetilde\eta_2^2 & \widetilde\eta_2\widetilde\eta_3 \\
\widetilde\eta_1\widetilde\eta_3 & \widetilde\eta_2\widetilde\eta_3 & \widetilde\eta_3^2\end{array}\right)}_{\textrm{leading order}}+c_1 
\underbrace{\left(\begin{array}{ccc}
\widetilde\xi_1^2 & \widetilde\xi_1\widetilde\xi_2 & \widetilde\xi_1\widetilde\xi_3 \\
\widetilde\xi_1\widetilde\xi_2 & \widetilde\xi_2^2 & \widetilde\xi_2\widetilde\xi_3 \\
\widetilde\xi_1\widetilde\xi_3 & \widetilde\xi_2\widetilde\xi_3 & \widetilde\xi_3^2\end{array}\right)}_{\textrm{perturbation}}
\right].
\label{eq:neu_per}
\end{eqnarray}
where $c_0\equiv (1+\tan^2\theta_W)/2$ and $c_1\equiv (1+\cot^2\theta_W)/2$ for the second case of eq.~(\ref{eq:check}). This mass matrix consists of leading $\sim \mathcal{O}(\tilde{\eta_i} \tilde{\eta_j})$ and sub-leading, i.e., perturbation $\sim \mathcal{O}(\tilde{\xi_i} \tilde{\xi_j})$ terms. The leading order matrix can generate only one massive light neutrino at the tree level as it is of rank one.  Using  the projective nature of the neutrino mass matrix one can rotate away one of the three mixing angles ($\theta_{12}$) and finally the non-zero eigenvalue and the two non-zero mixing angles turn out to be
\begin{eqnarray}
[m_{\nu}]_3 \simeq \frac{m_{3/2}}{16\pi^2}c_0|\vec{\widetilde\eta}|^2,~~~~~
\tan\theta_{13} \simeq -\frac{\widetilde\eta_1}{\left(\widetilde\eta_2^2+\widetilde\eta_3^2\right)^{1/2}},~~~~~
\tan\theta_{23} \simeq \frac{\widetilde\eta_2}{\widetilde\eta_3}.
\end{eqnarray}
Here, the heaviest neutrino mass generated from the leading order terms of the effective neutrino mass matrix is denoted as $[m_{\nu}]_3$. These two mixing angles   constrain the trilinear parameters $\widetilde\eta$. We now rewrite  eq.~(\ref{eq:neu_per})  in a compact form as
\begin{eqnarray}
(m_{\nu})_{ij} &\simeq & \frac{m_{3/2}}{16\pi^2}|\vec{\tilde{\eta}}|^2 c_0 \left(\begin{array}{ccc}
x\frac{\widetilde\xi_1\widetilde\xi_1}{|\vec{\widetilde\xi}|^2} & x\frac{\widetilde\xi_1\widetilde\xi_2}{|\vec{\widetilde\xi}|^2} & x\frac{\widetilde\xi_1\widetilde\xi_3}{|\vec{\widetilde\xi}|^2}\\
x\frac{\widetilde\xi_1\widetilde\xi_2}{|\vec{\widetilde\xi}|^2} & x\frac{\widetilde\xi_2\widetilde\xi_2}{|\vec{\widetilde\xi}|^2} & x\frac{\widetilde\xi_2\widetilde\xi_3}{|\vec{\widetilde\xi}|^2}\\
x\frac{\widetilde\xi_1\widetilde\xi_3}{|\vec{\widetilde\xi}|^2} & x\frac{\widetilde\xi_2\widetilde\xi_3}{|\vec{\widetilde\xi}|^2} & 1+x\frac{\widetilde\xi_3\widetilde\xi_3}{|\vec{\widetilde\xi}|^2}\end{array}\right),
\end{eqnarray}
where $x=c_1 |\vec{\widetilde\xi}|^2/c_0 |\vec{\widetilde\eta}|^2$. Under the assumption $\xi_3\sim 0$, the heaviest neutrino mass remains almost unaltered, and the other masses and the third mixing angle get generated as
\begin{eqnarray}
[m_\nu]_1 = 0,~~~~~ 
[m_{\nu}]_2 \simeq \frac{m_{3/2}}{16\pi^2}c_1\left(\widetilde\xi_1^2+\widetilde\xi_2^2\right),~~~~~
\tan \theta_{12} \simeq \frac{\widetilde\xi_1}{\widetilde\xi_2}. 
\end{eqnarray}
The recent  data for the neutrino mass and mixing angles are shown in table~\ref{Table:nufit}~\cite{Esteban:2016qun}.
\begin{table}[h!]
\centering
 \begin{tabular}{||c| c c||} 
 \hline
 Parameters & Normal ordering & Inverted ordering \\ [0.5ex] 
 \hline\hline
 $\sin^2\theta_{12}$ & $0.306^{+0.012}_{-0.012}$ & $0.306^{+0.012}_{-0.012}$ \\ 
 \hline
 $\sin^2\theta_{23}$ & $0.441^{+0.027}_{-0.021}$ & $0.587^{+0.020}_{-0.024}$ \\
 \hline
 $\sin^2\theta_{13}$ & $0.0216^{+0.00075}_{-0.00075}$ & $0.0218^{+0.00076}_{-0.00076}$\\
 \hline
 $\frac{\Delta m_{21}^2}{10^{-5}~\text{eV}^2}$ & $7.50^{+0.19}_{-0.17}$ & $7.50^{+0.19}_{-0.17}$ \\
 \hline
 $\frac{\Delta m_{3\ell}^2}{10^{-3}~\text{eV}^2}$ & $2.524^{+0.039}_{-0.040}$ & $-2.514^{+0.038}_{-0.041}$ \\ [1ex] 
 \hline
\end{tabular}
\caption{Latest values of the neutrino oscillation parameters within $3\sigma$~\cite{Esteban:2016qun} error range. Note that, $\ell=1,2$ for normal and inverted hierarchies respectively.}
\label{Table:nufit}
\end{table}
Using these fitted parameters we find the following constraints for normal (NH) and inverted (IH) hierarchies respectively
\begin{eqnarray}
&& m_{3/2} \widetilde\eta_3^2 \simeq 6.5\times 10^{-3}~\text{keV~(NH)},~~~~~~ m_{3/2} \widetilde\eta_3^2 \simeq 7.1\times 10^{-3}~\text{keV~(IH)},\nonumber \\
&& m_{3/2} \widetilde\xi_2^2 \simeq 4.5\times 10^{-4}~\text{keV~(NH)},~~~~~~ m_{3/2} \widetilde\xi_2^2 \simeq 2.6\times 10^{-3}~\text{keV~(IH)},\nonumber \\
&& \widetilde\eta_1 \simeq -0.2 \widetilde\eta_3~\text{(NH)},\hspace{3cm} \widetilde\eta_1 \simeq -0.2 \widetilde\eta_3~\text{(IH)},\nonumber \\
&& \widetilde\eta_2 \simeq 0.89 \widetilde\eta_3~\text{(NH)},\hspace{3.1cm}\widetilde\eta_2 \simeq 1.19 \widetilde\eta_3~\text{(IH)}, \nonumber \\
&& \widetilde\xi_1 \simeq 0.67 \widetilde\xi_2~\text{(NH)},\hspace{3.2cm}\widetilde\xi_1 \simeq 0.67 \widetilde\xi_2~\text{(IH)}.
\label{eq:numasscons}
\end{eqnarray}
Moreover, a legitimate bound on the gravitino mass can be obtained when the DM constraints are taken into consideration. For example, for a MeV scale gravitino, $\widetilde\eta\sim 10^{-3}$ and $\widetilde\xi\sim 10^{-4}$ are the ball-park numbers which satisfy neutrino mass constraints. 

Before proceeding further, we would like to briefly sketch the distinct features of our methodology compared to that described in earlier papers. In Ref.~\cite{U1R1,U1R2},  the idea of generating neutrino masses and mixing was proposed in a framework with leptonic $R$-symmetry and pseudo-Dirac gauginos. Two separate cases of $R$-symmetry breaking were considered namely anomaly mediation (AMRB) and Planck mediation (PMRB). In the AMRB scenario, it was noted that the neutrinos remain massless at the tree-level and only become massive after radiative corrections. This is noticeably different from our case. In our paper, we confine ourselves in the AMRB scenario as the Planck mediated $R$-breaking operators can be sequestered away. Unlike~\cite{U1R2}, neutrino masses and mixings can be explained at the tree level itself because of our specific $R$-charge assignments and the presence of trilinear terms ($\xi_i \widehat S\widehat H_u\widehat L_i$ and $\eta_i \widehat H_u.\widehat T.\widehat L_i$). In addition, the $R$-charge of the $\widehat{L}$ superfield was considered to be zero in \cite{U1R2} which allowed sneutrinos to acquire large VEVs. Such VEVs were also not constrained by the Majorana masses of the neutrinos. Hence, sneutrinos could also play the role of a down-type Higgs field. Following our charge assignment, sneutrino VEVs can be rotated away completely. Hence, noticeably different phenomenology and signatures at the collider are obtained. Moreover, in Ref.~\cite{U1R2} a tentative bound on the gravitino mass of $m_{3/2}<0.5$~GeV was obtained from the study of neutrino masses. However, such a constrain is not be applicable in our scenario as the neutrino mass is proportional to $m_{3/2}\xi^2 (m_{3/2}\eta^2)/16\pi^2$.  The superpotential couplings $\xi$, $\eta$ give an additional handle to fit neutrino masses and hence the strict bound on the gravitino mass can be somewhat ameliorated. Further, the bound on the gravitino mass in our case comes from DM constraints, which in turn gives an estimate on the model parameters.
\subsection{Charged fermion sector}
The presence of trilinear terms ($\eta_i\widehat L_i\widehat T\widehat H_u$) in the superpotential results in a mixing between charged leptons and charginos. As the sneutrino VEVs are rotated away, therefore the gauge couplings between the charged leptons and bino/wino vanishes. However, the mixing between higgsino like charginos and charged leptons is present. Such a mixing may lead to lepton number violating (LNV) processes. Therefore, a robust bound on the relevant superpotential parameters can be obtained from LNV studies. The chargino mass matrix in the basis $(e_L,\mu_L,\tau_L,\widetilde W^{-}, \widetilde T^{-}, \widetilde R_d^{-}, \widetilde H_d^{-})$ and $(e_R,\mu_R,\tau_R,\widetilde W^{+}, \widetilde T^{+}, \widetilde R_u^{+}, \widetilde H_u^{+})$ can be written as:
\begin{eqnarray}
m_{\widetilde f^+} &=& \left(\begin{array}{ccccccc}
v_d y_{11} & v_d y_{21} & v_d y_{13} & 0 & \frac{v_u\widetilde\eta_1}{\sqrt{2}} & 0 & v_S\widetilde\xi_1-v_T\widetilde\eta_1 \\
v_d y_{12} & v_d y_{22} & v_d y_{23} & 0 & \frac{v_u\widetilde\eta_2}{\sqrt{2}} & 0 & v_S\widetilde\xi_2-v_T\widetilde\eta_2 \\
v_d y_{13} & v_d y_{23} & v_d y_{33} & 0 & \frac{v_u\widetilde\eta_3}{\sqrt{2}} & 0 & v_S\widetilde\xi_3-v_T\widetilde\eta_3 \\
0 & 0 & 0 & M_2 & M_2^D-g v_T & 0 & g v_u\\
0 & 0 & 0 & M_2^D + g v_T & M_T & -\lambda_T^d v_d & 0 \\
0 & 0 & 0 & 0 & \lambda_T^u v_u & 0 & -\mu_u \\
0 & 0 & 0 & g v_d & 0 & -\mu_u & 0\\
\end{array}\right).
\end{eqnarray}
We  note that for  $\widetilde\eta\sim 10^{-3}$, the mixing elements and the masses of the charged leptons remain unaltered. The constraint appearing from lepton flavor violation~\cite{Bartoszek:2014mya,Gavela:2009cd} can be translated into the following bound $\left(v^2/M_D^2\right)~\widetilde\eta^2 < 10^{-5}\implies \widetilde\eta < 5\times 10^{-3}$. 
\section{\label{sec:5}Gravitino as dark matter}
In order to check the viability of gravitino DM, we first discuss the production of gravitino and then its possible decay modes. The lifetime of the gravitino should be more than the age of the universe for it to become a feasible DM candidate. In addition, the decaying DM has to be consistent with constraints from diffuse gamma rays also.

\subsection{Production of gravitinos}
\label{subsec:7}
 To start with, we consider the evolution of the universe dictated by the standard model of cosmology. This assumes the presence of an inflationary phase after the Big-bang. Any trace of pre-inflationary matter or radiation gets diluted because of the expansion and super cooling of the universe. The inflationary phase continues till the inflaton field reaches the minima of the scalar potential. 

The total amount of energy stored in the inflaton then gets transformed into relativistic matter leading to drastic rise in the temperature and entropy of the universe. As a result, the  universe  reaches its maximum temperature known as the reheating temperature $T_R$. The gravitino can then reach thermal equilibrium with its environment in the post-reheating period. Although the number density of  gravitino was negligible to start with, it gets generated through scattering and decays of particles (squarks and gluinos) which are in thermal equilibrium within the plasma. Assuming $m_{\text{SUSY}}\ll T_R$ (at the computational level), the thermal relic density of the gravitino~\cite{Bolz:2000fu,Pradler:2006qh} can be estimated as

\begin{eqnarray}
\Omega_{3/2} h^2 &\sim& 0.1 \left(\frac{1~\text{GeV}}{m_{3/2}}\right)\left(\frac{T_R}{10^7~\text{GeV}}\right)\left(\frac{m_{\widetilde g}}{2~\text{TeV}}\right)^2,
\label{gvdm}
\end{eqnarray}
where $\Omega_{3/2} h^2\sim 0.1199$~\cite{Ade:2013zuv}. 
The gravitino exchanges energy and momentum  with the particles already present in the thermal bath. This leads to a state of maximum entropy in which the distribution function follows Fermi-Dirac or Bose-Einstein statistics $f(p)=\left[{\text{exp}\left(\frac{E(p)-\mu}{T}\right)\pm 1}\right]^{-1}$, where $\mu$ is the chemical potential and `+' (-) sign stands for fermions (bosons).
The thermal production of gravitinos would require $T_R \gg m_{\widetilde g}\sim 2$ TeV~\footnote{In our scenario, gluinos are very heavy compared to the electro-weak gauginos. Thus a pair of gluinos may decay to $q\bar q\widetilde\chi_1^0$, where $q$'s are the first two generation quarks. Based on this type of decay, LHC provides stringent constraints on the gluino mass $\geq$ 2 TeV~\cite{ATLAS:2016kts,CMS:2016mwj}.} otherwise the production of gravitinos would be exponentially (Boltzmann) suppressed as: $\text{exp}\left(-m/T\right)$~\cite{Bomark:2014yja}. Even though this outcome heavily depends on the exact SUSY spectrum, the constraint on the reheating temperature can be mostly satisfied with
\begin{eqnarray}
m_{3/2} \gtrsim 200~\text{keV}.
\label{eq:gravitino_mass}
\end{eqnarray}
Such a gravitino would constitute a cold DM~\cite{Fujii:2003nr}. 
Gravitino `freeze-in' can also play an important role in the fast decay of the superpartners in the thermal equilibrium (the short lifetime of gluinos, squarks and sleptons induce this process). In such  scenarios the gravitino DM may suffer from an over-abundance problem. The only way to circumvent this issue would require lowering the reheating temperature below the SUSY scale~\cite{Baltz:2001rq}. But such a low reheating temperature ($T_R\sim 2$~TeV) could be troublesome~\cite{Fujii:2002fv} from the perspective of thermal leptogenesis where typical values of the reheating temperature is required to be around $\sim 10^8-10^9$ GeV.

\subsection{Gravitino decay width and life-time}
\label{subsec:6}
As mentioned earlier, gravitinos are metastable in our framework with typically large life-time. The reason is two fold: first the couplings are suppressed by the supersymmetry breaking scale  and second the smallness of the superpotential couplings $\eta$ and $\xi$. In general a small photino-neutrino mixing allows the  gravitino to decay into a photon and a neutrino. At tree level this decay channel is prohibited in our case as sneutrino VEVs are rotated away. However, for small gravitino mass this decay is feasible at one-loop and turns out to be the dominating one over the three body decay of gravitino into fermions. Thus we can safely ignore the consequences of the latter decay mode and  assume that our DM candidate, i.e., gravitino decays into a photon and a neutrino producing two monochromatic lines at energy exactly equal to $m_{\text{DM}}/2$. This decay width~\cite{Takayama:2000uz} of gravitino is well approximated by
\begin{eqnarray}
\Gamma (\widetilde G\rightarrow\nu_i\gamma) &\simeq& \frac{1}{32\pi}|U_{\gamma\nu}|^2 \frac{m^3_{3/2}}{M_P^2}, 
\label{gravitino_decay}
\end{eqnarray}
where $\left|U_{\gamma\nu}\right|^2=\sum_{a=i+4}\left|\cos\theta_W Z_{a1}+\sin\theta_W Z_{a2}\right|^2$ and $M_P\sim 2.4\times 10^{18}$ GeV is the reduced Planck mass. We approximate $|U_{\gamma\nu}|^2\simeq \sum_{a=i+4}|N_{a1}|^2$ where $m=1,2,3$. The gravitino decay width can be further simplified in terms of the model parameters as
\begin{eqnarray}
\Gamma(\widetilde G\rightarrow \nu_i\gamma) &\simeq& \frac{3}{32\pi}\left[\frac{\xi_i v_u}{M_D}\right]^2 \frac{m^3_{3/2}}{M_P^2}\cos^2\theta_W.
\label{decay_simp}
\end{eqnarray}
Consequently, the lifetime of the gravitino turns out to be 
\begin{eqnarray}
\tau_{3/2} \simeq \frac{32\pi}{3\cos^2\theta_W}\left(\frac{M_D}{\widetilde\xi_i v_u}\right)^2\frac{M_P^2}{m_{3/2}^3}~\text{GeV}^{-1}. 
\label{eq:lifetime}
\end{eqnarray}
In order for the gravitino to become a valid DM candidate, the first priority is its lifetime should be greater than the age of the universe, which is around $4.32\times 10^{17}$ sec. From eq.~(\ref{eq:lifetime}) it is straightforward to find that for $\widetilde\xi\sim 10^{-3}$
\begin{eqnarray}
m_{3/2}\lesssim 10~\text{GeV}.
\end{eqnarray}

\subsection{Photon flux from gravitino decays}
\label{subsec:8}
\begin{figure}[b!]
	\center
		\includegraphics[width=12cm, height=9cm]{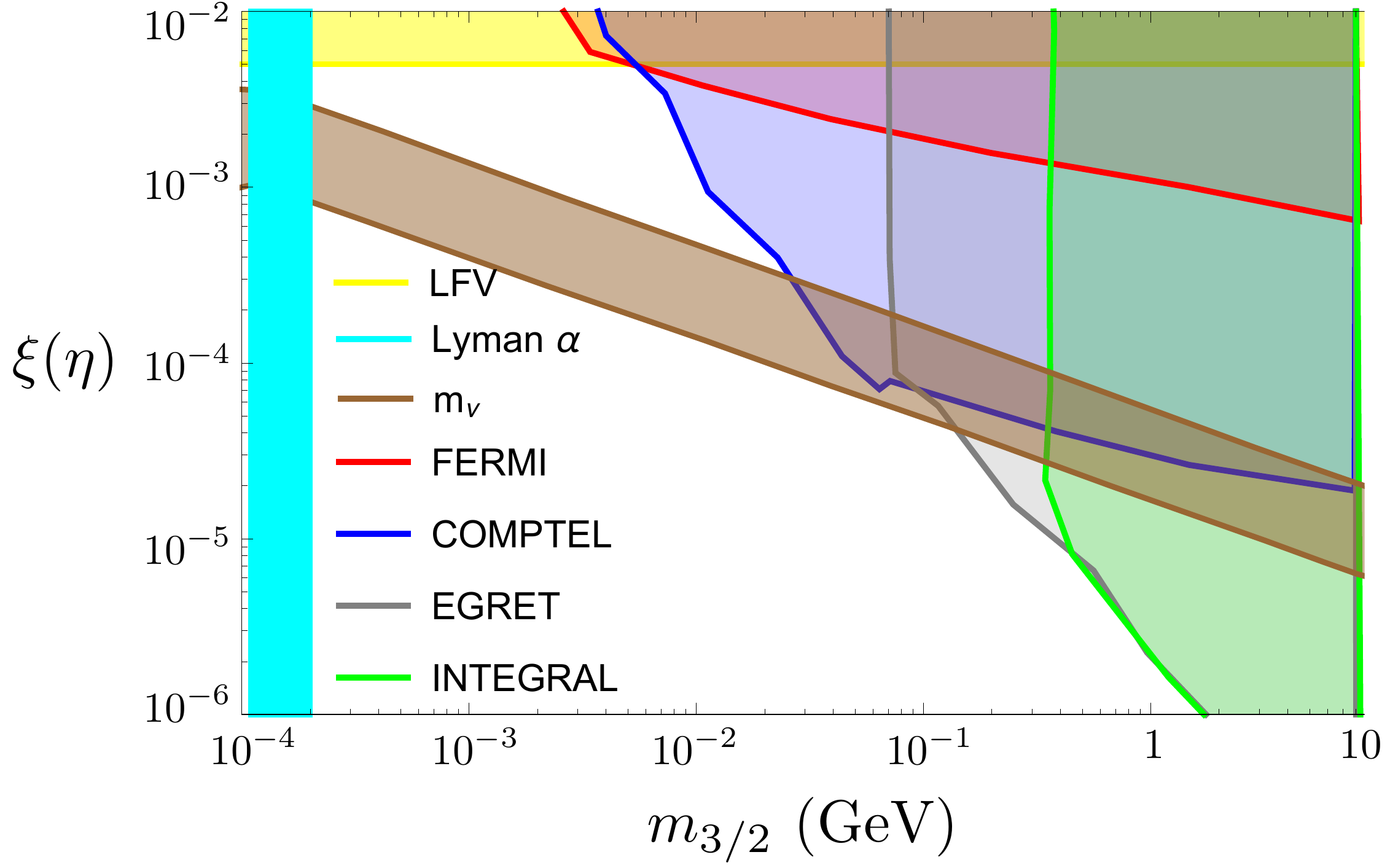}
	\caption{We show the allowed parameter space of the model. The photon fluxes considered from FERMI (red), COMPTEL (blue), EGRET (gray), INTEGRAL (green) experiments are translated to the $\xi (\eta)-m_{3/2}$ plane. Constraints from the Lyman-$\alpha$ forest experiment rules out gravitino mass less than 200 keV. We also overlay the neutrino mass constraints from eq.~(\ref{eq:numasscons}) where the brown shaded region satisfies constraints from neutrino mass data. The yellow shaded region is ruled out from the lepton flavor violating decays.}
	\label{fig:boat1}
\end{figure}

To circumvent the stringent constraints from the diffuse gamma ray sources, the lifetime of the DM needs  to be much greater~\cite{Essig:2013goa} than the age of the Universe ( $\gtrsim\;10^{24}-10^{28}$ sec.). The neutrino flux however is vanquished by the atmospheric neutrino background in the energy range of few MeV to GeV which makes its detection more difficult compared to the gamma ray flux. In general one can think of two typical sources for a diffuse gamma ray background. Firstly, due to DM decay in the nearby Milky way halo and secondly DM decay at cosmological distances. For inner galaxy constraints, data from INTEGRAL~\cite{Bouchet:2008rp} or COMPTEL~\cite{Weidenspointner:2000aq} are used to probe photons within the mass range of 20 keV -- 2 MeV. While the EGRET~\cite{Sreekumar:1997un}  and FERMI~\cite{Ackermann:2012pya} experiments probe a region of 20 MeV -- 10 GeV and 200 MeV -- 10 GeV diffuse photons respectively. In most of the  cases Navarro--Frenk--White (NFW)~\cite{Navarro:1995iw,Navarro:1996gj} profile of DM density  is used. The bounds on the energy of the photons are quite sensitive to the DM density profile and may vary up to $\mathcal O(20\%)$. 

In fig.~\ref{fig:boat1}, we show the allowed parameter space  in the $m_{3/2}-\xi (\eta)$ plane compatible with the existing experimental constraints. The photon fluxes are taken from FERMI (red), COMPTEL (blue), EGRET (gray), INTEGRAL (green) experiments. Noticeably,  Lyman-$\alpha$ forest experiment rules out gravitino of mass less than 200 keV. We overlay the neutrino mass constraints depicted by the brown-shaded allowed region, obtained from eq.~(\ref{eq:numasscons}). The yellow region is disfavored from lepton flavor violating decays. From fig.~\ref{fig:boat1}, it is clear that the allowed range for gravitino mass lies in between
\begin{eqnarray}
200~\text{keV}\lesssim m_{3/2}\lesssim 0.1~\text{GeV}.
\end{eqnarray}
Notice, the more robust upper bound on the gravitino mass comes from Lyman-$\alpha$ forest experiment. This limit coincides with the bound assuming the reheating temperature above the scale of superpartner masses. The lower limit on gravitino mass is achieved using the constraint from diffuse gamma ray fluxes in conjunction with neutrino oscillation data. Using eq.~(\ref{gvdm}) for $m_{\widetilde g}\equiv2$~TeV, we can obtain a corresponding bound on the reheating temperature
\begin{eqnarray}
2\times 10^3~\text{GeV}\lesssim T_R\lesssim 10^6~\text{GeV}.
\end{eqnarray}
In this regime, electroweak baryogenesis~\cite{Fischler:1990gn,Campbell:1990fa,Dreiner:1992vm,Kuzmin:1985mm} can explain the baryon asymmetry of the universe as the reheating temperature is well above the electroweak phase transition temperature. Also, in presence of R-symmetry, there are no $A-$terms in the scalar potential.  This may allow the presence of large CP-violation originating from complex mass terms (Dirac gaugino masses, $\mu_{u/d}$ etc.) without affecting flavor, electric dipole moment and other low energy constraints \cite{Kumar:2011np, Kribs:2007ac, Fok:2012fb}. This large CP-violation may turn out to be a suitable source of CP-asymmetry which can be translated into baryon asymmetry~\cite{Kribs:2007ac} through sphaleron effects. Another viable option for baryogenesis could be the Affleck-Dine mechanism~\cite{Affleck:1984fy}. During or after the reheating of the universe, the scalar superpartners carrying baryon or lepton numbers would decay into the SM fermions. The net baryon number carried by the SM fermions then may explain the observed excess of baryons over anti-baryons.

We would also like to point out that the viability of gravitino DM in $U(1)_R$ models have already been discussed  in~\cite{U1R1}. In that framework, the alternative assignment of $R$-charges and large sneutrino VEVs ensure the tree level decay of the gravitino into a photon and a neutrino to be the most prominent. Such a scenario is severely constrained from  diffuse gamma ray searches and can be tackled with much diluted gravitino density \cite{U1R1}. This can be achieved by assuming the reheating temperature to be lower than the SUSY scale. As a result, gravitino cannot explain the observed relic abundance of the universe. In our case, however, the absence of such tree level decay mode makes the gravitino a viable DM candidate and also the reheating temperature can be relatively larger compared to that in~\cite{U1R1}.
\section{\label{sec:6}Collider Phenomenology}
In our model,  gravitino turns out to be the LSP.  With suitable choice of parameters, we can choose a valid SUSY spectrum with lightest neutralino to be the next-to-minimal supersymmetric particle (NLSP). Thus we would be left with a scenario where all the supersymmetric particles decay to the lightest neutralino which  decays to a gravitino accompanied by either photon / $Z$-boson / Higgs. Such interactions are suppressed by the Planck scale and the resulting decay width will be very small, i.e., corresponding lifetime would be too large for the decay to occur within the collider. In addition, the NLSP also undergoes $R$-parity violating decay modes primarily in the following channel
\begin{eqnarray}
\widetilde\chi_{1,2}^0 &\rightarrow& h~\nu_i,\; \;  \gamma~\nu_i. 
\end{eqnarray}
The dominant decay width is noted down as~\cite{Bobrovskyi:2012dc}
\begin{eqnarray}
\Gamma(\widetilde\chi_{i}^0\rightarrow h\nu_m) &=& \frac{\alpha m_{\widetilde\chi_i^0}}{16\sin^2\theta_W}\left|\frac{\widetilde\xi_m\cos\alpha}{\sqrt{2}}N_{i3}N_{11}\right|^2\left[1-\frac{m_h^2}{m^2_{\widetilde\chi_i^0}}\right]^2.
\end{eqnarray}
Here we assume that $\widetilde\chi_{1,2}^0$ is either bino or singlino type. Notice that the same final state topology would arise if the neutralino is wino or triplino type. In that case the relevant parameter would be $\widetilde\eta$ instead of $\widetilde{\xi}$. This particular decay mode of the light neutralino to a neutrino and a Higgs boson gives rise to an interesting di-Higgs signature at the colliders~\cite{Biswas:2016ffy}. Moreover, a pair produced charginos can also decay to pair of Higgs boson associated with opposite sign charged leptons through the $\widetilde\eta$ coupling. This is a distinct feature of our scenario. For our case, the typical values of $\widetilde\xi$ and $\widetilde\eta$ are around $\mathcal O(10^{-3})$. This leads to rather prompt decays at the colliders. This is also different from the standard RPV case where characteristic signals come from longer decay lengths and displaced vertices due to the smallness of the sneutrino VEVs. In the wino-higgsino decoupled scenario the lightest neutralino is primarily an admixture of the bino-singlino  states. In addition to the $h\nu$ final state,  $Z\nu$ and $W^{\pm}\ell^{\mp}$ decay modes are also present. One can easily have light higgsinos in this framework without much modifications in the neutrino sector and would also have interesting collider implications~\cite{Mukhopadhyaya:1998xj}.

In generic bilinear RPV models, the mixing between the Higgs and the sneutrino induces a slight mass splitting between sneutrinos and anti-sneutrinos which gives rise to sneutrino
oscillation~\cite{Chun:2001mm} signatures. This mixing is controlled by the $B_{\alpha}H_u\widetilde L_i$ term in the scalar potential. Also the sneutrino VEVs generate trilinear couplings involving lepton (quark) and slepton (squark) fields which are interesting channels to look for~\cite{Roy:1996bua}. However, such signals are not probable in our case as the bilinear terms are rotated away completely. Moreover, the presence of small $R$-breaking effects would create a slight mass splitting between the pair of Dirac neutralinos. These pseudo-Dirac neutralinos can give rise to neutralino oscillation~\cite{Grossman:2012nn} signatures at the colliders. In typical cases, these neutralinos can decay to a $h\widetilde G$. The difference between the distribution of the displaced vertices between the almost degenerate pseudo-Dirac neutralinos are a smoking gun signature of such a framework. However, in our case, the primary decay modes are $h\nu_i$, $Z\nu$ and $W^{\pm}\ell^{\mp}$ and the decay to gravitino is largely suppressed. In fact, a detailed study of the trilinear terms makes our scenario phenomenologically distinct, rich and explorable at the LHC.

\section{\label{sec:7}Conclusion}

Neutrinos and dark matter play a very important role in understanding the physics beyond the SM.  Supersymmetric models while solving the hierarchy problem can also address the issues pertaining to neutrino masses and dark matter.  However, the present searches by both ATLAS and CMS have found no significant excess in their pursuit of superpartners. As a result, stringent constraints are obtained on the superpartner masses. In this light, models with $R$-symmetry and Dirac gauginos are well motivated as they can relax these constraints. Therefore,  in this paper, we come up with an $R$-symmetric  SUSY scenario with specific $R$-charges leading to bilinear and trilinear ``$R$-parity violating" terms at the superpotential. Our prime aim in this paper is to explain how active light neutrino masses and mixing can be generated.  In the process we also discuss  the generation of a Higgs mass around 125 GeV.  Then we  motivate the requirement of $R$-symmetry breaking through anomaly mediation. The bilinear terms from the superpotential and the scalar potential can be rotated away simultaneously due to the suitable choice of basis. However,  the trilinear terms will always be there playing an important role in generating neutrino masses and mixing. We constraint the relevant superpotential parameters while fitting neutrino masses for both normal and inverted hierarchies.  In our scenario neutrino masses are generated at the tree level itself which is vastly different than the standard RPV-MSSM scenarios. In standard RPV-MSSM scenario, lepton number violation can emerge from  the bilinear ($\epsilon_i \widehat H_u\widehat L_i$) as well as the trilinear ($\lambda_{ijk}\widehat L_i\widehat L_j\widehat E_K^c$, $\lambda^{\prime}_{ijk}\widehat L_i\widehat Q_j\widehat D_K^c$) terms in the superpotential. In case of bilinear RPV, only one of the neutrinos acquire mass at the tree level, while the other two become massive through one-loop induced effects. But such a process necessitates significant tuning between the model parameters. Again, in presence of  trilinear RPV, all the neutrinos acquire masses at the one-loop level. As these operators are constrained from lepton flavor violating (LFV) processes, a minuscule room is left to fit neutrino masses and mixing after satisfying all other constraints. Unlike usual scenarios, in our framework, neutrino masses and mixing can be explained at the tree level itself even with not so fine-tuned values of the superpotential parameters $\xi$, $\eta$  $\sim \mathcal O(10^{-3}-10^{-4})$. In passing we would like to mention that such choice of parameters are also compatible with LFV constraints.

In our proposal, LSP gravitino mass turns out to be the order parameter of $R$-breaking and it qualifies to be an excellent DM candidate. We  explore this possibility by considering the production and decays of gravitino. While incorporating the constraints from diffuse gamma ray experiment, the model becomes more predictive leaving an allowed region of the gravitino mass in the range $200~\text{keV}\lesssim m_{3/2}\lesssim 0.1~\text{GeV}$. The collider signatures are also quite different from the standard RPV-MSSM case. In our framework, lightest neutralino decay leads to di-Higgs signatures. Similarly, pair-production of charginos may also lead to a pair of Higgs accompanied by  opposite signed leptons. These decays are controlled by the superpotential parameters $\xi$, $\eta$ and for their suitable values $\sim \mathcal O(10^{-3})$ prompt decay of electro-weakinos  could be observed.

\section*{Acknowledgements}
We thank Sourov Roy for many helpful discussions and insights. SC would also like to thank the hospitality of IIT Kanpur and HRI Allahabad where a part of the project was completed. 


\end{document}